\documentclass[twocolumn,amsmath,amssymb,superscriptaddress]{revtex4-1}

\usepackage{bm}
\usepackage[dvipdfmx]{graphicx}
\usepackage{braket}
\usepackage{bbm}
\usepackage[T1]{fontenc}
\usepackage{times}
\usepackage{xcolor}
\usepackage[colorlinks,allcolors=blue]{hyperref}
\usepackage{graphicx}
\usepackage{bm}
\usepackage{mathrsfs}
\usepackage{dcolumn}
\usepackage{ulem}
\usepackage{dsfont}
\usepackage{amsmath}
\usepackage{amsfonts}
\usepackage{amssymb}

\renewcommand{\vec}[1]{{\ensuremath{\bm{\mathrm{#1}}}}}

\renewcommand{\exp}[1]{\ensuremath{{\mathrm{e}^{#1}}}}

\begin{document}

\title{Magnon Spectrum of the Amorphous Ferromagnet Co$_4$P from Atomistic Spin Dynamics}

\author{Mai Kameda}
\affiliation{Institute for Materials Research, Tohoku University, Sendai 980-8577, Japan}
\affiliation{Department of Applied Physics, Nagoya University, Nagoya 464-8603, Japan}

\author{Gerrit E. W. Bauer}
\affiliation{Institute for Materials Research, Tohoku University, Sendai 980-8577, Japan}
\affiliation{WPI Advanced Institute for Materials Research, Tohoku University, Sendai 980-8577, Japan} 
\affiliation{Zernike Institute for Advanced Materials, University of Groningen, 9747 AG Groningen, The Netherlands}

\author{Joseph Barker}
\affiliation{Institute for Materials Research, Tohoku University, Sendai 980-8577, Japan}
\affiliation{School of Physics and Astronomy, University of Leeds, Leeds LS2 9JT, United Kingdom}
\affiliation{Bragg Centre for Materials Research, University of Leeds, Leeds LS2 9JT, United Kingdom}

\date{\today}

\begin{abstract}
An anomaly in the magnon dispersion of the amorphous ferromagnet Co$_{4}$P,
often referred to as a `roton-like' excitation, attracted much attention
half a century ago. With the current interest in heat and spin currents in
amorphous magnets, we apply modern simulation methods, combining reverse Monte
Carlo to build the atomic structure and the stochastic Landau-Lifshitz
equation for spin dynamics, to re-investigate the magnetic excitation spectrum. We find two
magnon valleys, one at the origin and another at a finite wavenumber close to the
observations, but without a magnon gap. We conclude that the second dip is due to
Umklapp scattering caused by residual long-range order, which may be an alternative
explanation of the putative roton excitation. Our study paves the way to study magnon 
transport in amorphous magnets and related spintronic applications.
\end{abstract}
\maketitle

{\it Introduction.} -- Amorphous magnets are technologically important due to their highly tuneable
coercivity and magnetisation for, e.g., power transformers and magnetic
memories. Commonly used materials include random rare-earth--transition metal alloys such as GdFeCo for
magnetooptical devices~\cite{Meiklejohn_ProcIEEE_74_1570_1986} and CoFe alloys for
spintronic devices~\cite{Parkin_NatMater_3_862_2004,
Jen_JApplPhys_99_053701_2006}. The phenomenology of these materials is often
not much different from ordered materials, displaying conventional ferro or
ferrimagnetic order. 
On the other hand, in thermally induced switching~\cite{Barker_2013_SciRep_3_1} or long-range magnon transport~\cite{Wesenberg_2017_NatPhys_13_10,Ochoa_PhysRevB_98_054424_2018} the local atomic arrangement appears to be important. Magnon transport in
amorphous systems is currently a topic of debate since experimental results
contradict each other~\cite{Gomez_Perez_ApplPhysLett_116_032401_2020,Yang_PhysRevB_104_144415_2021}.

In the 1970's, amorphous Co$_{4}$P attracted much
interest~\cite{Coey_JApplPhys_49_1646_1978, Handley-1987, kaneyoshiBook-1984},
because neutron scattering experiments discovered a local minimum in its
magnon dispersion at a finite wavenumber \cite{Mook_PhysRevLett_34_1029_1975}.
This feature is reminiscent of the dip in the phonon dispersion of liquid He
caused by the ``roton'' excitation that
limits superfluidity \cite{Feynman-1953, Feynman-1956}. We use this material
in the following work as a relatively simple ferromagnetic representative for
amorphous magnets \cite{Coey_JApplPhys_49_1646_1978}.

The simplest approach to compute the magnon dispersion in an amorphous alloy
is the quasi-crystalline approximation
(QCA)~\cite{1978_Kaneyoshi_JPhysSocJpn_45_6}. It is based on an
angle averaged approximation of the amorphous atomic structure, 
expressed by an atomic pair-correlation function.
The energy of a spin wave $\varepsilon$ with wave number $Q$ in an ensemble of local moments $\mu$ then reads
\begin{equation}
\varepsilon_{\mathrm{QCA}}(Q)=4\pi\mu \rho_{\mathrm{m}} \int J(r_{ij})g(r_{ij})\left(
1-\frac{\sin Qr_{ij}}{Qr_{ij}}\right)  r_{ij}^{2}dr_{ij}, \label{eq:qca}%
\end{equation}
where $\rho_{\mathrm{m}}$ is mean density of the magnetic atoms, $J(r_{ij})$ is the exchange
interaction dependent on distance $r_{ij}=|\vec{r}_{j}-\vec{r}_{i}|$, where
$\vec{r}_{i}$ ($\vec{r}_{j}$) denotes the position of the $i$-th ($j$-th) magnetic atom, and
$g(r_{ij})$ is the pair-correlation function. With physically motivated models
for $J(r_{ij})$ and $g(r_{ij}),$ the QCA predicts spectra that can be a useful
guide for small wave numbers, including a dip close to the wave numbers of the
roton-like feature. However, this minimum is much shallower than was observed
in Co$_{4}$P. 
Higher-order corrections deepen the valley a little~\cite{Roth-1975, Roth-1976}. 
Numerical simulation also showed local dips but were severely limited by the available
computing power and were based on linear spin wave theory~\cite{Alben-1976}. 
Therefore the suspicion lingers that the magnetic roton feature is an experimental
artefact~\cite{1982_Shirane_PhysRevB_26_5}. Motivated by the intriguing
observation of enhanced spin transport in amorphous
materials~\cite{Wesenberg_2017_NatPhys_13_10} and by the greatly increased
computational power in the past decades, we revisit the problem of the
non-monotonous spin wave dispersion in amorphous Co$_{4}$P.

Our simulations expose a magnon dispersion and neutron scattering cross
section that is an intriguing mix of the QCA predictions and a remnant of
crystal symmetry. Instead of a minimum in the magnon dispersion, we predict a
mirror image of the $Q=0$ magnons with parabolic dispersion and a narrow
linewidth at the $Q$ vector of the historical neutron scattering measurements, 
close to the Brillouin zone boundary of a virtual crystal.
We conclude that residual Umklapp scattering causes magnon dispersion minima at large wave numbers
\cite{1982_Shirane_PhysRevB_26_5}.

{\it Methods.} -- The atomic positions in amorphous alloys are not precisely known, 
but they are not distributed completely randomly either. Our task is to find statistical
ensembles that on average describe the specific material properties. The
observed roton-like gap depends sensitively on, for example, the alloy
composition~\cite{Mook_PhysRevLett_34_1029_1975, Mook-1977}, so it appears to
have a structural origin. Here we generate the atomic positions of the
amorphous alloy by the reverse Monte Carlo (RMC) method~\cite{2001_McGreevy_JPhysCondensMatter_13_46} under constraints of established observations, which should produce a more physically realistic model than
building an alloy by random packing~\cite{Scott-1962, Cargill-1970,
Gaskell-1979, Finney-2013}.

We employ the RMC++ code~\cite{Gereben2007_JOptoelectronAdvMater_9} in order
to profit from the experimental X-ray, neutron and polarised neutron
diffraction data for Co$_{4}$P~\cite{Sadoc_MaterSciEng_23_187_1976}. We start
with an FCC lattice with substitutional disorder in the form of randomly
distributed Co and P in a 4:1 ratio. In each iteration step (a) two atoms can
be swapped or (b) a single atom can be moved a small distance in a random
direction~\cite{Gereben2007_JOptoelectronAdvMater_9}. The volume is kept
constant due to periodic boundary conditions. The mean-square cost function
for a scattering function ($i=$X-ray, neutron, polarised neutron)
reads $\chi_{i}^{2}=\frac{1}{\sigma_{i}^{2}}\sum_{Q}\left(  \mathcal{S}%
_{i}^{\mathrm{calc}}(Q)-\mathcal{S}_{i}^{\mathrm{exp}}(Q)\right)^{2}$,
where $\sigma_{i}$ is a weight that reflects the confidence level of a data
set, $\mathcal{S}_{i}^{\mathrm{calc}}(Q)$ and $\mathcal{S}_{i}^{\mathrm{exp}%
}(Q)$ are the calculated and measured scattering functions for a discrete
set of scattering vectors. Each move that lowers the total cost function
$\chi^{2}=\sum_{i}\chi_{i}^{2}$ is accepted unconditionally while those that
increase $\chi^{2}$ are accepted with a probability of $\mathrm{exp}%
(\chi_{\mathrm{old}}^{2}-\chi_{\mathrm{new}}^{2})$, where $\chi_{\mathrm{old}%
}^{2}$ and $\chi_{\mathrm{new}}^{2}$ are the cost function values
before and after the move.

We model the atoms by hard spheres with radii $r_{\mathrm{Co}}=1.25$~\AA {}
and $r_{\mathrm{P}}=1.00$~\AA , ignoring the chemical bonding.
We implement the known feature of amorphous compounds like Co$_{4}$P that the
anions (P in this case) almost never
touch~\cite{Sadoc_MaterSciEng_23_187_1976} by an increased cost 
for the P atoms closer than $2.75\,$\AA .

\begin{figure}[ptb]
\includegraphics[width=\linewidth]{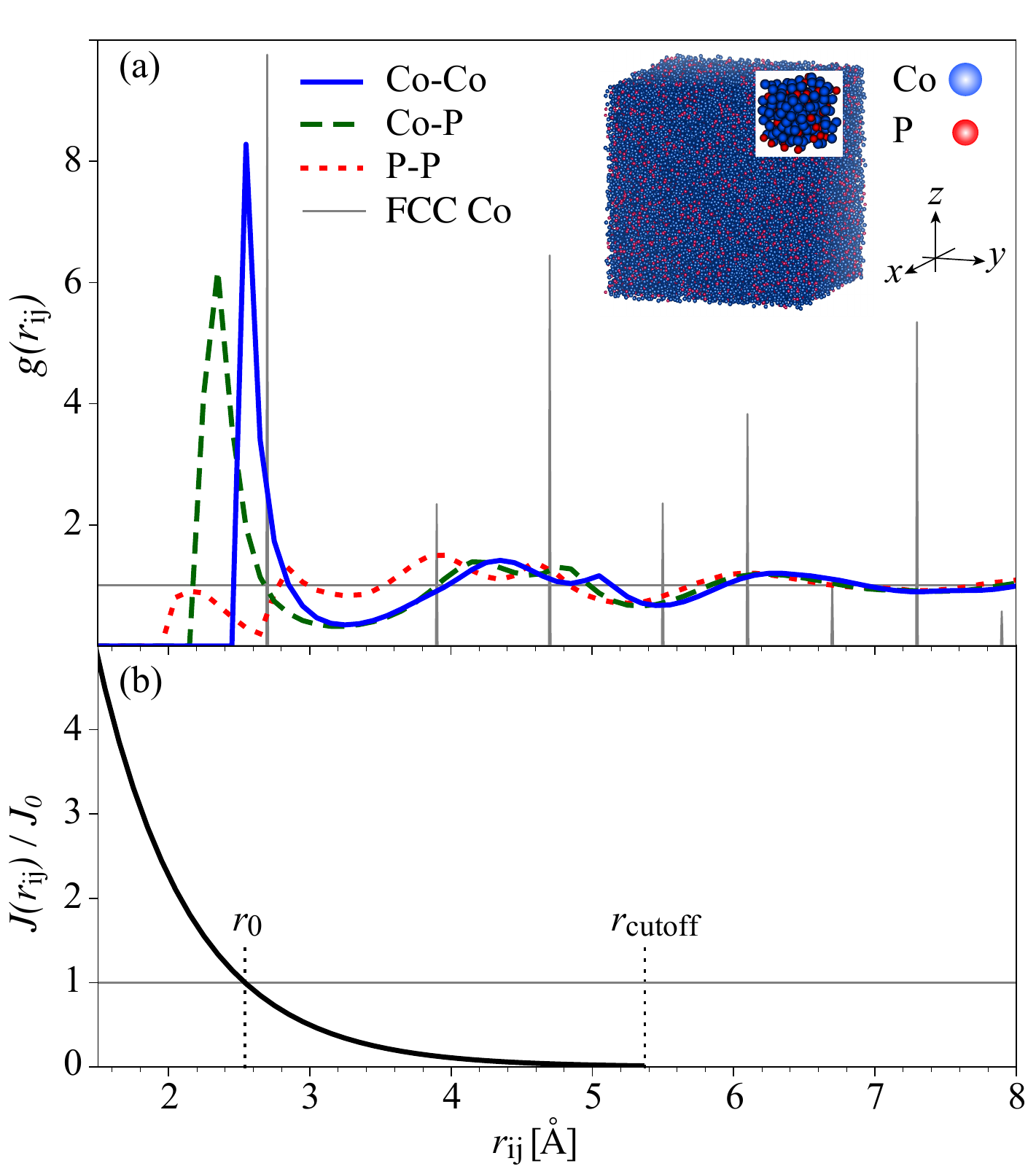}
\caption{(a) Correlation functions $g(r_{ij})$ for Co-Co, Co-P, and P-P pairs in
Co$_{4}$P generated by RMC. The vertical lnes illustrate the $\delta$-function
correlations in an crystalline FCC system with the same volume. The inset shows example of an RMC generated amorphous Co$_{4}$P with 62500 atoms (blue = Co and red = P).
(b) Single-exponential exchange interaction $J(r_{ij})$ used in the atomistic spin
simulations, which we set to zero for $r_{ij}>r_{\mathrm{cutoff}}=5.45\,$\AA .
}%
\label{fig:rmc_pair_correlation}%
\end{figure}

After the Monte-Carlo iterations converged to a minimum of the cost function, as shown in Fig.~\ref{fig:rmc_pair_correlation}(a), 
we compute the magnetic properties by atomistic spin dynamics
(ASD)~\cite{Gyorffy_JPhysFMetPhys_15_1337_1985, Skubic-2008}. The $i$-th Co
atom at $\vec{r}_{i}$ has a local moment $\mu={\mu_{\mathrm{B}}}$ (Bohr magneton)~\cite{Mook_PhysRevLett_34_1029_1975} and direction $\vec{S}(\vec{r}_{i})$ with $|\vec{S}|=1$. The non-magnetic P atoms are treated as 
vacancies~\cite{Durand-1976}. Assuming that anisotropies and superexchange
interactions average out in random alloys, we adopt the isotropic Heisenberg
model Hamiltonian,
\begin{equation}
\mathscr{H}=-\tfrac{1}{2}\sum_{i\neq j}J({r}_{ij})\vec{S}(\vec{r}_{i}%
)\cdot\vec{S}(\vec{r}_{j})-\mu\vec{B}\cdot\sum_{i}\vec{S}(\vec{r}_{i}),
\label{eq:Hamiltonian}%
\end{equation}
where $\vec{B}$ is an external magnetic field. The exchange interaction
$J(r_{ij})$ extends beyond nearest neighbours. Our knowledge of the exact
functional form of the exchange has not progressed much in the past decades,
so we implemented several options, such as a step function, exponential decay,
and oscillating (RKKY) functions~\cite{kaneyoshiBook-1984} and with different
ranges. Since the results do not change significantly, we concluded that the precise distance
dependence is not an important issue. In the following, we use the exponential
decay, shown in Fig.~\ref{fig:rmc_pair_correlation}(b),
\begin{equation}
J(r_{ij})=J_{0}{\mathrm{exp}}{\left(  -\frac{r_{ij}-r_{0}}{w}\right)  }\text{
for }r_{ij}>r_{0},
\end{equation}
where $J_{0}=6.733$~meV, $r_{0}=2.54$~\AA , and a decay length $w=0.66$~\AA .
With these values the curvature of the magnon dispersion $\epsilon(Q)$
corresponds to the experimental spin wave stiffness of amorphous Co$_{4}$P,
$D=\frac{1}{2}[\partial^{2}\epsilon(Q)/\partial Q^{2}]_{Q=0}=135\,$%
meV\AA $^{2}$~\cite{kaneyoshiBook-1984}. Truncating the exchange at large
distances by setting $J(r_{ij})=0$ for $r_{ij}>5.45$~\AA {} does not affect
the results, but reduces the computational cost. In order to emphasize the
effects of disorder, we compare results for amorphous Co$_{4}$P with those for
hypothetical crystalline FCC cobalt with the same volume and parameters.

The Landau-Lifshitz equation for a local moment reads
\begin{equation}
\frac{d\vec{S}(\vec{r}_{i})}{dt}=-\gamma\left[  \vec{S}(\vec{r}_{i})\times
\vec{H}(\vec{r}_{i})+\alpha\vec{S}(\vec{r}_{i})\times\left(  \vec{S}(\vec
{r}_{i})\times\vec{H}(\vec{r}_{i})\right)  \right]
\end{equation}
where $t$ is time, $\gamma=1.76\times10^{11}$~$\mathrm{rad}\ \mathrm{s^{-1}T^{-1}}$ is the
gyromagnetic ratio, $\alpha=0.01$ is a damping constant, and $\vec{H}(\vec
{r}_{i})=\vec{\xi}(\vec{r}_{i})-(1/\mu)(\partial\mathscr{H}/\partial\vec
{S}(\vec{r}_{i}))$ is the effective magnetic field on the spin at $\vec{r}_i$. $\vec{\xi}(\vec{r}_{i})$ is
a fluctuating field that provides a temperature to the spin system. We use a
quantum thermostat that obeys the fluctuation-dissipation
theorem~\cite{Barker_2019_PhysRevB_100_14},
\begin{equation}
\langle\xi_{a}(\vec{r}_{i},t)\rangle=0;\ \langle\xi_{a}(\vec{r}_{i})\xi
_{b}(\vec{r}_{j})\rangle_{\omega}=\delta_{ij}\delta_{ab}\frac{2\alpha
}{\gamma\mu\beta}\frac{\hbar\omega}{\exp{\beta\hbar\omega}-1},
\end{equation}
where $a$ and $b$ are Cartesian components, $\omega$ is the frequency,
$\beta=(k_{\mathrm{B}}T)^{-1}$ is the inverse thermal energy with $T$ is temperature, 
$\hbar$ is
Planck's constant, $\langle\cdots\rangle\ $is a statistical time average,
and $\langle\cdots\rangle_{\omega}$ is a statistical average in frequency space. This thermostat
describes thermodynamic properties well up to the Curie temperature
~\cite{Ito_2019_PhysRevB_100_6,Barker_2020_ElectronStruct_2_4}. 
The combination of RMC for the atomic structure, the ASD with the quantum
thermostat, and the computational power to handle large systems all
drastically improve previous approaches to simulate random magnets~\cite{Alben-1976, Roth-1975, Roth-1976, 1978_Kaneyoshi_JPhysSocJpn_45_6}.

Our algorithm first equilibrates a 
large number of spins (62500) to a constant temperature. 
After reaching the steady state, we carry out the thermodynamic
averaging of the desired properties by collecting fluctuating spin
trajectories around their equilibrium values up to 0.4~ns.
Their time averages lead to the thermodynamic properties, while
the power spectra are Fourier transforms of the space-time spin-spin
correlation functions. Even though the systems size is already large, we
confirm ergodicity by averaging over 10 realisations of the amorphous
arrangement of atoms.

{\it Results.} -- Fig.~\ref{fig:rmc_pair_correlation}(a) shows the calculated pair-correlation
functions $g_{mn}(r_{ij})=n_{mn}(r_{ij})/(4\pi r_{ij}^{2}\Delta r_{ij}\rho
_{n})$ of Co-Co, Co-P, and P-P pairs in amorphous Co$_{4}$P and in crystalline
FCC Co-Co, where $n_{mn}$ is the number of neighbours of atomic type $n$ at
distance from $r_{ij}$ to $r_{ij}+\Delta r_{ij}$ from an atom of type $m$.
$\Delta r_{ij}$ is a binning width of a histogram and $\rho_{n}$ is the number
density of atoms of type $n$. $g_{\mathrm{PP}}$ is featureless with a weak
maximum at $\sim4$~\AA , so P is nearly homogeneously distributed and only few
P atoms touch each other, as intended by the extra cost for their proximity.
The observed double peaked behaviour in $g_{\mathrm{CoCo}}$ around $4.4$~\AA {}
and $5.0$~\AA {} indicates short-range order, a common feature of
amorphous metalloids~\cite{kaneyoshiBook-1984}. The average number of nearest
neighbours, counted as atoms within a radius $r_{\mathrm{nbr}}$, is 7.53 for Co-Co ($r_{\mathrm{nbr}}=3.1\,\mathrm{\mathring{A}}$), 
1.96 for Co-P ($r_{\mathrm{nbr}}=3.0\,\mathrm{\mathring{A}}$), 
and 0.30 for P-P ($r_{\mathrm{nbr}}=2.75\,\mathrm{\mathring{A}}$), 
where $r_{\mathrm{nbr}}$ has been chosen based on the first peak of $g(r_{ij})$ for each pair. 
We show an example of an RMC generated atomic configuration  in the inset to Fig~\ref{fig:rmc_pair_correlation}(a).

\begin{figure}[ptb]
\includegraphics[width=\linewidth]{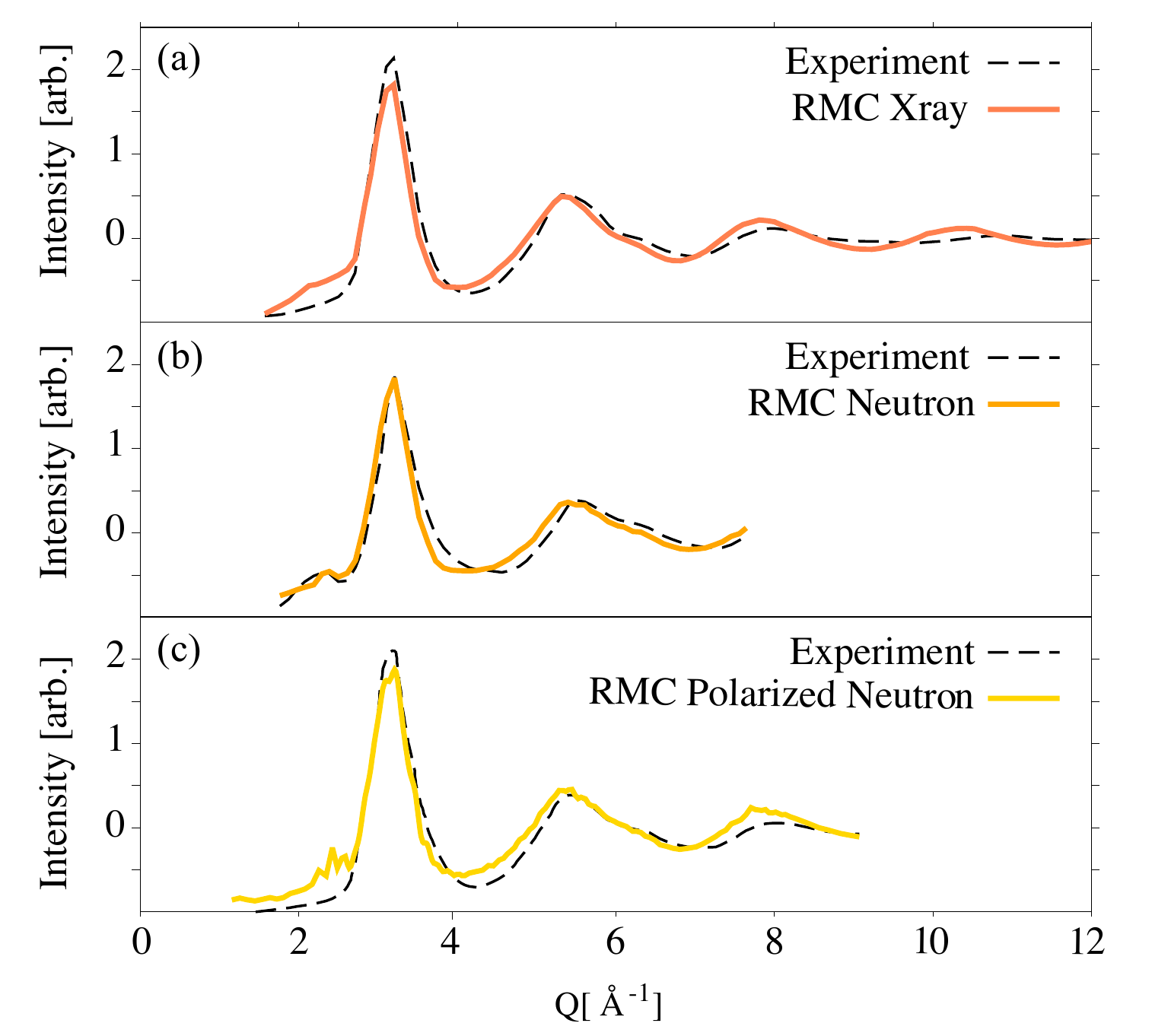} \caption{(a) X-ray, (b) neutron,
and (c) polarized neutron scattering functions of Co$_{4}$P. The solid lines
are the results of the RMC simulations. The dashed lines are adopted from the
experiments \cite{Sadoc_MaterSciEng_23_187_1976}. }%
\label{fig:rmc_scattering}%
\end{figure}

By design, the pair correlation functions agree well with those
inferred from the scattering functions~\cite{Sadoc_MaterSciEng_23_187_1976},
as demonstrated in Fig.~\ref{fig:rmc_scattering}(a)-(c).

\begin{figure}[ptb]
\includegraphics[width=\linewidth]{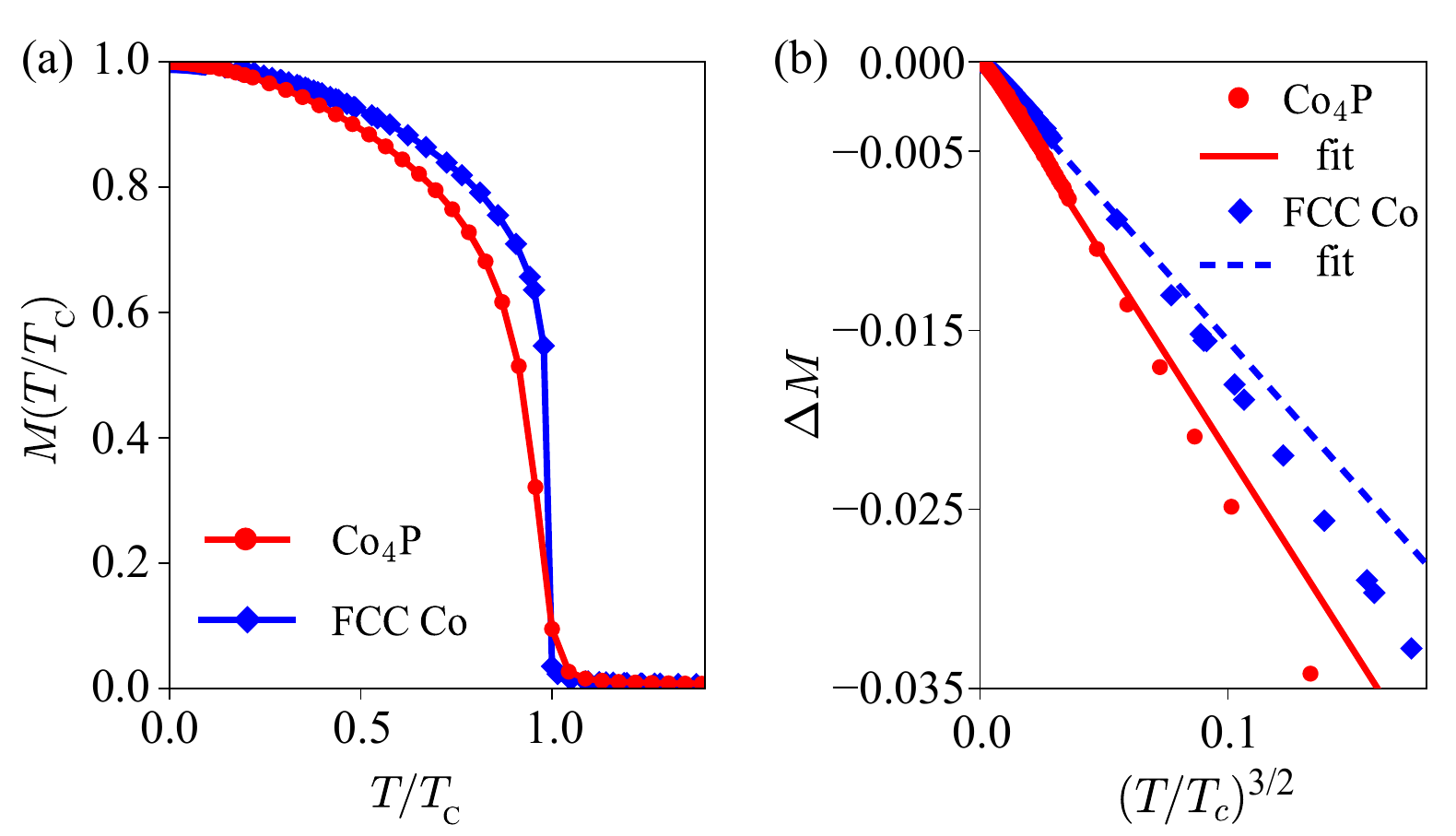}
\caption{(a) Calculated temperature dependence of the magnetisation of
amorphous Co$_{4}$P and hypothetical crystalline FCC Co. $T_{\mathrm{C}}\sim500\,$K for
both systems. (b) Normalised temperature $(T/T_{\mathrm{C}})^{3/2}$ vs. reduced
magnetisation $\Delta M=M(T/T_{\mathrm{C}})-1$. Solid and dashed lines are
low-temperature fits to Bloch's law $\Delta M=-B_{3/2}(T/T_{\mathrm{C}})^{3/2}$. }%
\label{fig:magnetisation}%
\end{figure}

Fig.~\ref{fig:magnetisation} shows the temperature dependence of the
dimensionless magnetisation $\vec{M}(T)=\langle(1/N)\sum_{i=1}^{N}\vec{S}(\vec{r}_{i})\rangle_{T}$, where $N$ is total number of magnetic atoms, for the amorphous and crystalline systems. 
The latter has a larger lattice constant than the physical FCC Co, which is a good
metal with $s$-$d$ hybridized bands and high Curie temperature $T_{\mathrm{C}}$. 
The susceptibilities (not shown) of both the hypothetical FCC Co and Co$_{4}$P peak at $T_{\mathrm{C}}\sim500\,\mathrm{K}$.
The experimental $T_{\mathrm{C}}$ of Co$_{4}$P is
620-720K~\cite{Cochrane_PhysRevLett_32_476_1974,
Mook_PhysRevLett_34_1029_1975} so even though we reproduce the experimental
spin wave stiffness, the calculated $T_{\mathrm{C}}$ is lower than observed.
At low temperatures ($T\ll T_{\mathrm{C}}$) the
magnetisation of both crystalline and amorphous systems decreases according to
Bloch's law $M(T)=1-B_{3/2}(T/T_{\mathrm{C}})^{3/2}$. We find $B_{3/2}=0.16$ for the
FCC Co, which is very close to the experimental value of $B_{3/2}=0.17$ for
FCC lattices~\cite{kaneyoshiBook-1984}. 
The larger $B_{3/2}=0.22$ for amorphous Co$_{4}$P reflects a reduced spin wave
stiffness~\cite{kaneyoshiBook-1984}. 
However, it is about a half the reported $B_{3/2}\sim0.45$~inferred from 
magnetometry measurements~\cite{Cochrane_PhysRevLett_32_476_1974}. 
These discrepancies of $T_{\mathrm{C}}$ and $B_{3/2}$ might be due to non-collinearities in the magnetic ground state caused by the superexchange via P or local anisotropies~\cite{Continentino_JPhysFMetPhys_9_L145_1979}. Moreover, the value of $B_{3/2}$ inferred from experimental neutron scattering measurements of the stiffness has generally been smaller than from magnetometry for a variety of amorphous ferromagnets~\cite{Axe-1975, Axe-1977, Continentino_JPhysFMetPhys_9_L145_1979}. 
It is a large parameter space to explore and we do not pursue the issue in more detail here.

Next we address the unusual roton-like dip observed in the inelastic neutron
scattering spectra of Co$_{4}$P. To this end we compute the inelastic neutron scattering cross section, 
\begin{align}
&\mathcal{S}(\vec{Q},\omega)  =\frac{g_{n}^{2}r_{0}^{2}}{2\pi\hbar}%
f^{2}(Q)\sum_{ab}\left(  \delta_{ab}-\hat{Q}_{a}\hat{Q}_{b}\right)  \sum
_{i,j}e^{-i{\vec{Q}\cdot\vec{r}_{ij}}}\nonumber\\
&  \times\int_{-\infty}^{\infty}e^{-i{\omega t}}\left[  \left\langle
S_{a}(\vec{r}_{i},0)S_{b}(\vec{r}_{j},t)\right\rangle -\left\langle S_{a}%
(\vec{r}_{i})\right\rangle \left\langle S_{b}(\vec{r}_{j})\right\rangle
\right]  dt, 
\label{eq:scattering_equation}%
\end{align}
where $g_{n}=1.931$ is the neutron g-factor,
$r_{0}=e^{2}/m_{e}c^{2}=2.8\,$fm is the classical electron radius with $e$,
$m_{e}$, and $c$ the elementary charge, the electron mass, and the speed of light, respectively, $f(Q)$ is the atomic form factor of
Co~\cite{crystal_tables}, $\vec{Q}$ is the scattering vector, and $\hat{\vec{Q}}=\vec{Q}/|\vec{Q}|$. The spin-spin
correlation function in Eq.~\eqref{eq:scattering_equation} cannot be expressed
analytically for amorphous magnets, even in linear spin wave theory. We
compute the correlation function from the spatiotemporal dynamics of our large
spin cluster without linearization, thereby including the magnon-magnon
interactions to all orders.

\begin{figure}[ptb]
\includegraphics[width=\linewidth]{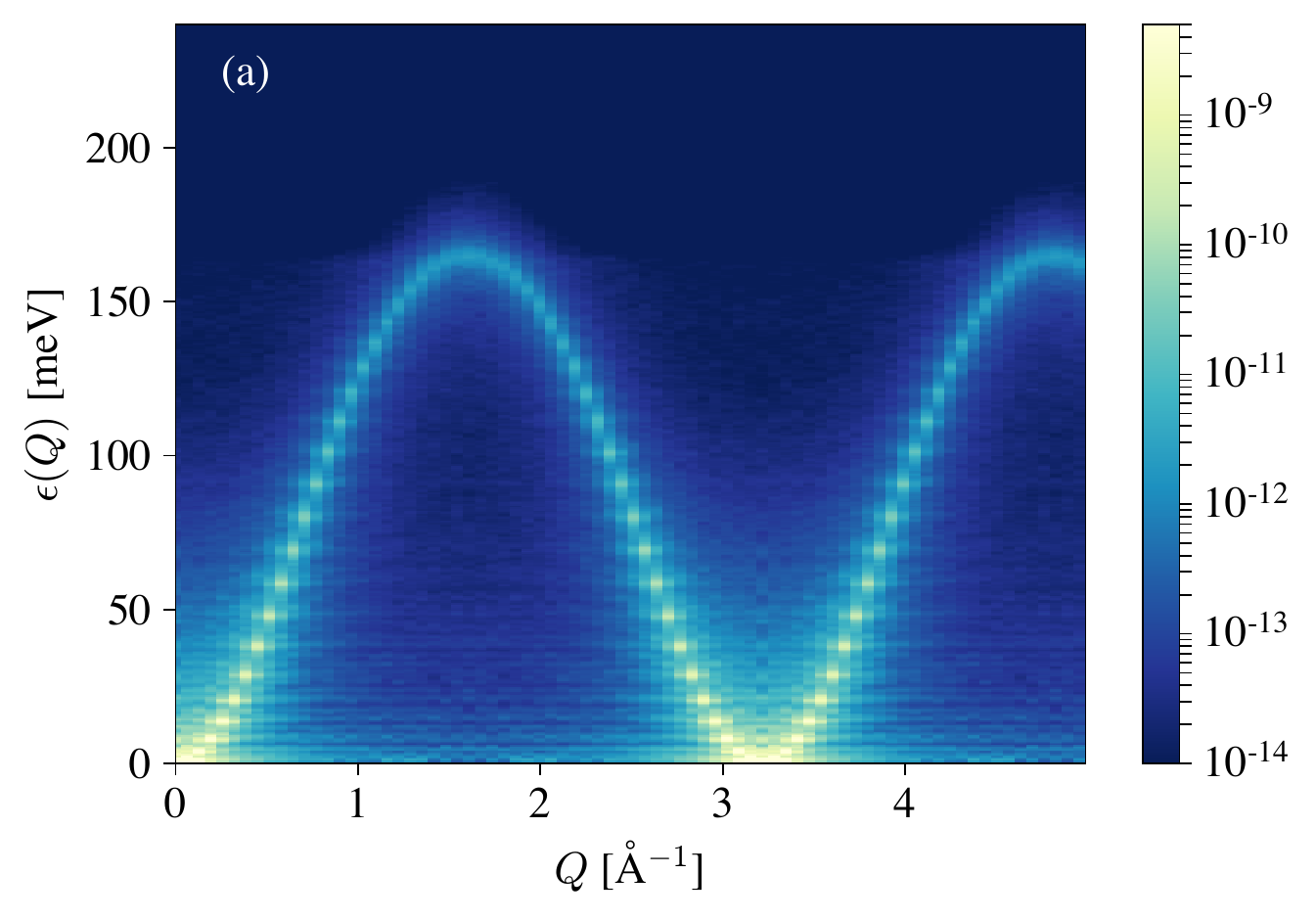}
\includegraphics[width=\linewidth]{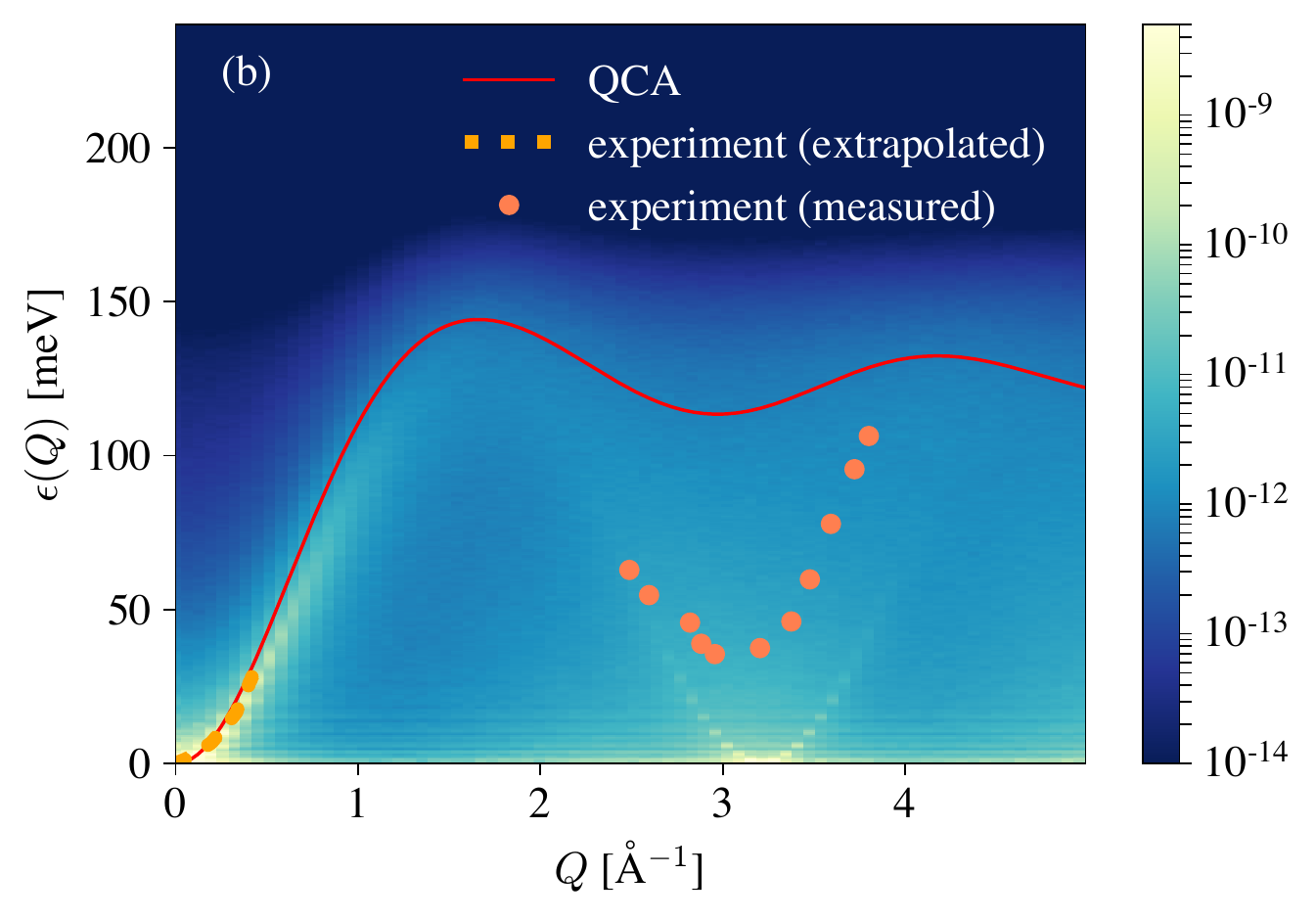} 
\caption{(a)-(b) Calculated
inelastic neutron scattering cross section Eq.~\eqref{eq:scattering_equation}
of (hypothetical) crystalline FCC Co (a) and Co$_{4}$P and the QCA analytic prediction (red solid line) (b) at temperature $T=300\,\mathrm{K}$.
In panel (b), the magnon (dashed orange line) and roton-like excitations (orange dots)~\cite{Mook_PhysRevLett_34_1029_1975} observed at room temperature are overlaid for comparison.}%
\label{fig:spectrum_qca_comparison}%
\end{figure}

In Fig.~\ref{fig:spectrum_qca_comparison}(a) and (b), we show the calculated
spectra for the crystalline model and amorphous Co$_{4}$P at room
temperature for $\vec{Q}\parallel [001]$. The excitations of the crystalline system are a periodic function
of momentum transfer in the extended Brillouin zone scheme, with an amplitude
that decays only weakly by the Co form factor. We observe a single magnon band, as expected for one spin per primitive unit cell.

We extract the spin wave stiffness $D$ from our spectra by a fit to $\epsilon(Q)=DQ^{2}$, for $Q<0.6$~\r{A}{}$^{-1}$. In the FCC model
[Fig.~\ref{fig:spectrum_qca_comparison}(a)] the stiffness is $D_{\mathrm{Co}%
}=182\mathrm{\,meV\,\mathring{A}}${}$^{2}$, whereas amorphous Co$_{4}$P model
[Fig.~\ref{fig:spectrum_qca_comparison}(b)] the spin waves are softer with
$D_{\mathrm{Co_{4}P}}=129\mathrm{\,meV\,\mathring{A}}${}$^{2}$, very close to
the experimental results. In crystalline magnets the magnon linewidth scales
as $\Gamma\sim\alpha\omega$, while the peaks are much broader in the amorphous
material, as expected in disordered systems. At high energies, $\epsilon
>50$~meV, the spectrum becomes diffuse, i.e. well-defined magnon excitations
cease to exist. In Fig.~\ref{fig:spectrum_qca_comparison}(b), we compare the
numerical results with Eq.~\eqref{eq:qca} in the QCA, which agrees quite well
close to the origin and appears to model the modulation of the diffuse
background at high energies.

At larger scattering vectors the amorphous magnetic spectrum shows a
clear feature with parabolic dispersion and zero gap, centred at $Q\approx
3.1$\thinspace\r{A}$^{-1}$, close to 
the first peak in the static structure factor (see
Fig.~\ref{fig:rmc_scattering}), but larger than the minimum of the shallow dip
predicted by the QCA. 
The calculated line width is close to that at the origin, indicating a coherent rather than diffuse magnon. The second minimum agrees with the reciprocal lattice vector of the FCC lattice with the same moment density, which was the starting configuration of the Monte-Carlo procedure. 
We observe analogous minima at the Brillouin zone boundary in other crystal directions as well such as $\vec{Q}\parallel [111]$ (not shown). 
However, in contrast to the crystalline system of the
artificial FCC Co where the spectrum repeats due to Bloch's theorem, the dip does not re-appear in the amorphous spectrum at higher values of $Q$. 
These minima are therefore caused by Umklapp scattering from residual periodicity on the scale of the magnon mean free path, as suggested
previously~\cite{1982_Shirane_PhysRevB_26_5}. But we cannot confirm that these
lead to a finite gap that is crucial for an exotic roton feature.

The thermodynamic properties are integrals in reciprocal space frequency and
momenta. In the FCC structure, these are limited to the crystal
momentum in the first Brillouin zone, but over all momenta for the amorphous
structure. A roton minimum with a finite gap at $\sim30$~meV should
affect the magnetisation at higher temperatures, but such deviations from 
Bloch's law have not been reported. The zero-gap dispersion feature at
$Q\approx3.1$~\r{A}$^{-1}$ is nearly identical to that at the origin and
contributes to the magnetisation without changes in the temperature scaling.

{\it Conclusion.} -- Our calculations of the spin wave spectrum of amorphous Co$_{4}$P find a
replica of the dispersion around the $\Gamma$-point at wavenumbers that
roughly agree with the `roton-like' dip observed by neutron scattering, but do
not reproduce the finite magnon gap. At higher energies, the spectrum is very
broad indicating strong scattering and the complete absence of coherent magnons.
At low frequencies, the spectra of amorphous Co$_{4}$P looks surprisingly
similar to that of crystalline ferromagnets. The sharp low frequency feature
in the second Brillouin zone implies a contribution in the magnon wave
functions that are coherently periodic over many lattice constants. In other
words, in spite of our efforts to generate an amorphous material based on the
experimental structure factors, the resulting structure retains some ordering.
We note that in the original neutron scattering experiments~\cite{Mook_PhysRevLett_34_1029_1975} there is a comment that the peaks in the Fourier transformed pair correlation function were sharper than usually seen in amorphous materials, hinting at the possibility that these samples also retained some short range order.
We hope that our work will inspire renewed experimental efforts to find out whether the roton gap is real. If the gap does not survive scrutiny, we have a
powerful method at hand to characterise the degree of disorder in non-ideal
amorphous magnets. Our calculations are also a good start for studying spin
transport properties in amorphous magnets for example by applying the Kubo formula~\cite{Mook-2017}.

We would like to thank K. Sato, K. Kobayashi, Y. Araki, and Y. Kawaguchi for
valuable discussions. This work was supported by the Graduate Program in
Spintronics (GP-Spin) at Tohoku University. Calculations were performed on
ARC4, part of the High Performance Computing facilities at the University of Leeds. G. E. W. B. was supported by JSPS KAKENHI (19H00645). 
J. B. acknowledges support from the Royal Society through a University Research Fellowship. M. K. was supported by Grant-in-Aid for JSPS Fellows
(JP19J20118) and GP-Spin at Tohoku University.

\bibliography{arXiv_Amorphous_Co4P}
\end{document}